\documentclass[10pt]{article}
\usepackage{latexsym}
\usepackage{amssymb}
\usepackage{amsfonts}
\usepackage{amsmath}
\usepackage{theorem}

\newtheorem{theorem}{Theorem}

\begin{document}

\title{Fluctuations in the Ising spin model \\ on a sparse random graph}

\author{Luca De Sanctis
\footnote{Department of Psychology, University of Bologna;
{\tt<luca.desanctis@gmail.com>}}}

\maketitle

\begin{abstract}
We compute the fluctuations of the magnetization and of the multi-overlaps
for the dilute mean field ferromagnet, in the high temperature region.
The rescaled magnetization tends to a centered Gaussian variable 
with variance diverging at the critical line. The rescaled multi-overlaps also 
tend to centered independent Gaussian variables, but their covariance remain
finite at the critical line.
\end{abstract}



\section{Introduction}

The mean field dilute ferromagnet is a very interesting model. 
It has direct connections with random graph theory, as its zero
temperature behavior reveals properties of the underlying random graph \cite{dsgu, dm}.
The dilute structure makes the model an intermediate one between the
finite dimensional Ising model and the fully connected Curie-Weiss model.
The ferromagnetic interactions simplify the search for basic results
(such as the computation of the free energy)
as compared to the spin glasses, but its physical behavior is still quite rich.
A complete control of the model is probably as difficult as in the case
of spin glasses and intriguing connections between ferromagnets and spin glasses
recently emerged, so that a general theory of these classes of spin models is now 
a fundamental open issue \cite{gparisi}.

Despite its importance, the model has been basically neglected. 
The powerful \emph{cavity} methods introduced by physicists could be easily guessed
to provide the correct free energy at all temperatures and other relevant
information, but no detailed study of the thermodynamics could be found in the 
physical literature.
The community of mathematicians instead mainly focused on pure graph theory or
on spin glasses, for which
numerous results from physicists were followed by major mathematical
breakthroughs \cite{broken, talabook}, while the 
dilute ferromagnet has been neglected.
Few years ago the interest of mathematical
physics, previously focused on mean field models, extended to dilute models,
but surprisingly only disordered systems were studied. 
The dilute ferromagnet gained some attention only very recently. A first study,
regarding the high temperature and the zero temperature behavior,
was based on interpolations and convexity methods \cite{dsgu},
in an attempt to complete the picture initiated with spin glasses \cite{ass},
then extended to dilute spin glasses \cite{lds1} and to the fully connected
ferromagnet as a test-case \cite{Houches, lds5} within a general ``structural''
approach. 
Surprisingly, the dilute mean field ferromagnet has not been fully framed 
so far in the context of these ``structural'' methods.
While the low temperature physical behavior of the model has not been 
well understood as yet, mainly due to the lack of physical literature about it, 
an important rigorous confirmation
of the cavity ansatz for the free energy at any temperature was recently published \cite{dm}.

Here we compute the fluctuations
of the rescaled magnetization (and of the multi-overlaps) in the high temperature region,
showing that such fluctuations diverge at the critical line (this holds for the magnetization
only, as the fluctuations of the rescaled multi-overlaps remain finite).


\section{The model, preliminary facts, previous results}

In this section we introduce the model and the notations,
provide some useful formulas which are at the basis of
almost all the calculations needed in this article, and report
previous results. More details can be found in \cite{dsgu}.

\subsection{Definitions}\label{definitions}

Given a set of $N$ points, the model is defined through
configurations $\sigma: i\to \pm 1$, $i=1,\dots , N$
of Ising spins. By $\{i_{\nu},j_{\nu},k_{\nu},l_{\nu}\}$, $\nu\in\mathbb{N}$, we will
denote families independent random variables all uniformly
distributed on $1,\ldots , N$.
The Hamiltonian is the random function defined by
\begin{equation}
\label{hamiltonian}
H_{N}(\sigma)=-\sum_{\nu=1}^{K}\sigma_{i_{\nu}}\sigma_{j_{\nu}}
\end{equation}
where $K$ is a Poisson random variable of mean $\alpha N$,
for some given $\alpha\in\mathbb{R}_{+}$ which is called
\emph{connectivity}.
The expectation with respect to the random choice of the spins
and with respect
to the Poisson random variable is denoted by $\mathbb{E}$,
and it is called \emph{quenched} expectation.
Given a non-negative real number $\beta$, whose physical interpretation
is the inverse of the temperature, the function
$$
A_{N}(\alpha,\beta)=\frac1N\mathbb{E}\ln\sum_{\sigma}
\exp(-\beta H_{N}(\sigma))
$$
is called pressure, and $-A_{N}(\alpha,\beta)/\beta$ is the free energy.
Given the simple relation between the two, we will indifferently use either
one or the other.
The sum
$$
Z_{N}(\beta)=\sum_{\sigma}
\exp(-\beta H_{N}(\sigma))
$$
is the (random) partition function, and the Boltzmann-Gibbs 
expectation of an observable $\mathcal{O}:\sigma\to 
\mathcal{O}(\sigma)\in\mathbb{R}$ is
$$
\Omega(\mathcal{O})=\frac{1}{Z_{N}}\sum_\sigma \exp(-\beta H_{N}(\sigma))
\mathcal{O}(\sigma)\ .
$$
When it is not confusing, we will omit the dependence
of $\Omega$ on $N$ or on the Poisson random variable appearing
in the Hamiltonian. When we omit the index $N$ in the pressure
we mean to have taken the thermodynamic limit: 
$A(\alpha, \beta)=\lim_{N\to\infty}A_{N}(\alpha,\beta)$.
The main physical quantity in this model is the magnetization of a configuration
\begin{equation}
\label{magnetization}
m(\sigma)=\frac1N \sum_{i=1}^{N}\sigma_{i}\ ,
\end{equation}
often simply denoted by $m$.
A further notation is $\langle \cdot \rangle = \mathbb{E}\Omega(\cdot)$.


Lastly, notice that according to our notations random variables with 
a subindex zero are independent of those appearing in the weights
$\Omega$ consists of, e.g. in the expression 
$\Omega(\sigma_{i_{0}}\sigma_{j_{0}})$.

\subsubsection{Useful formulas}

Given 
a generic Poisson variable $K$ with mean $\zeta$,
whose expectation is denoted by $\mathbb{E}$,
it is easy to verify that
\begin{equation}
\label{exponential}
\mathbb{E}a^{K}=e^{-\zeta(1-a)}
\end{equation}
for any positive real number $a$.

Another simple formula, $\mathbb{E}[g(K)]=\mathbb{E}[g(K+1)-g(K)]$
for a given function $g:\mathbb{N}\to\mathbb{R}$, is used to obtain
\begin{equation}
\label{alphaderivative}
\frac{\partial A_{N}(\alpha,\beta)}{\partial\alpha}=\ln\cosh\beta+
\mathbb{E}\ln[1+\Omega(\sigma_{i_{0}}\sigma_{j_{0}})\tanh\beta]\ .
\end{equation}

%



\subsection{The infinite connectivity limit}

Let us mention in what sense our model is a diluted version of the
Curie-Weiss one, and how the latter is recovered in a suitable
infinite connectivity limit. More details can be found in \cite{dsgu}.

Recall that the Hamiltonian of the Curie-Weiss (CW) model is 
$$
H^{(CW)}_{N}(\sigma)=\frac12 Nm^{2}(\sigma)\ ,
$$ 
and the associated pressure will be denoted by
$A_{N}^{(CW)}(\beta)$.
It is well known that if we let $\alpha\to\infty$, $\beta\to 0$ 
with $2\alpha\tanh\beta=\beta^{\prime}$
kept constant, the pressure of our model
tends to the one of the CW model, i.e. 
$A_{N}(\alpha,\beta)\to A^{(CW)}_{N}(\beta^{\prime})$
uniformly in the size of the system.

A simple proof can be obtained through interpolation \cite{dsgu}, considering,
for $t\in [0,1]$,
\begin{equation}
\label{interpolazione}
\hat{A}_{N}(t)\equiv\frac1N\mathbb{E}\ln\sum_{\sigma}\exp\bigg[
\beta\sum_{\nu=1}^{K_{1}}\sigma_{i_{\nu}}\sigma_{j_{\nu}}
+(1-t)\beta^{\prime} \frac12 Nm^{2}\bigg]\ ,
\end{equation}
$K_{1}$ being a Poisson random variable with mean $t\alpha N$.

\subsection{Symmetry breaking and dilution}

Let us define
$$
\tilde{A}_{S}(\alpha, \beta)=\ln 2+\alpha\cosh\beta\ .
$$
We know from \cite{dsgu} that if $2\alpha\tanh\beta\leq 1$ then
\begin{equation}
\label{HT}
A(\alpha,\beta)\equiv \lim_{N\to\infty}A_{N}(\alpha,\beta)=
\ln 2+\ln\cosh\beta=\tilde{A}_{S}(\alpha, \beta)\ .
\end{equation}
Interpolation (\ref{interpolazione}) allows for another simple but fundamental
observation: applying the fundamental theorem of calculus to (\ref{interpolazione})
ones observes that the difference between the limiting pressure $A(\alpha, \beta)$ and 
its symmetric approximation $\tilde{A}_{S}(\alpha, \beta)$
decreases with the dilution. This means that such a difference is larger 
for the CW model than for our diluted version:
\begin{multline}
\label{confronto}
A_{N}(\alpha,\beta)-\tilde{A}_{S}(\alpha, \beta)=A_{N}(\alpha,\beta)-\ln 2-\ln\cosh\beta \\
\leq A^{(CW)}_{N}(\beta^{\prime})-\ln 2\equiv A^{(CW)}_{N}(\beta^{\prime})-
\tilde{A}^{(CW)}_{S}(\beta^{\prime})\ ,
\end{multline}
where we are always assuming the constraint $2\alpha\tanh\beta=\beta^{\prime}$.
This is another way the crucial difference between ferromagnetic and 
glassy models reveals itself. In fact, in the case of symmetric distribution of the interactions
the odd terms disappear in the expansion of the pressure \cite{lds1}, and the first correction
to the zeroth term (the one without spin contribution, corresponding to the high temperature
symmetric pressure) is quadratic and negative \cite{lds1}. This makes the true pressure
smaller than its symmetric expression and the dependence on the order parameters 
(the multi-overlaps) convex. 
Moreover, the dilution increases the difference between the pressure and its 
symmetric expression \cite{gt}, hence when the symmetry is broken 
in the fully connected model
it is broken in the dilute model too (we are assuming the proper temperature rescaling). 
In the case of ferromagnetic interactions, in the expansion of (\ref{alphaderivative})
the odd terms contribute \cite{dsgu},
the magnetization being the first, and with positive sign; so the convexity
just mentioned is replaced here by the concavity in (\ref{alphaderivative}) with respect to
the order parameter $\Omega(\sigma_{i_{0}}\sigma_{j_{0}})$. This means that the pressure
gets larger when the interactions begin contributing, and it is thus larger than its
symmetric counterpart. Moreover, as explained by (\ref{confronto})
the dilution decreases the difference between the pressure and its 
symmetric expression, hence the occurrence of symmetry breaking in the fully
connected model does not imply the breaking of the symmetry in the dilute one.
So the idea is to study the model at temperature zero, where it is simpler, and
a symmetry breaking would imply the same transition at all temperatures (we again 
intend to keep $2\alpha\tanh\beta=\beta^{\prime}$ constant). 
This is carried out in \cite{dsgu}.

\subsection{The transition}

We here summarize some results obtained in \cite{dsgu}, adding a few comments.
The main observation is that our dilute model is in some sense delimited 
by its zero temperature limit and its infinite connectivity limit, both fully controlled
\cite{dsgu,Houches}. We are always assuming to move along the lines
$2\alpha\tanh\beta=\beta^{\prime}$. 
We are about to see that the two bounds squeeze at the critical line $\beta^{\prime}=1$.

As anticipated in the previous subsection, the study of the criticality of our model
cannot take advantage of the study of the criticality of the fully connected one only,
contrarily to the case of spin glasses.
It is necessary control what happens when the dilution is decreased, 
rather that increased, and as a consequence the temperature is decreased as well.
Consider 
\begin{equation*}
\label{}
A_{N}^{\prime}(\alpha,\beta)\equiv A_{N}(\alpha,\beta)-\tilde{A}_{S}(\alpha, \beta)
=A_{N}(\alpha,\beta)-\ln 2-\alpha\ln\cosh\beta\ .
\end{equation*}
It is proven in \cite{dsgu} that
$$
\liminf_{N\to\infty}A^{\prime}_{N}(\alpha,\beta)>0\qquad\mbox{if}\quad
2\alpha\tanh\beta>1\ .
$$
The idea of the proof is to keep $2\alpha\tanh\beta=\beta^{\prime}$ constant, so that
obviously $2(\tanh\beta) d\alpha+2\alpha(1-\tanh^{2}\beta)d\beta=0$, and to study
$dA^{\prime}(\alpha,\beta(\alpha))/d\alpha$ which turns out to be
non-negative increasing and convex in $\Omega(\sigma_{i_{0}}\sigma_{j_{0}})\tanh\beta$.
Such a quantity $\Omega(\sigma_{i_{0}}\sigma_{j_{0}})$ 
measures the difference between the pressure and its high
temperature symmetric expression, and its quenched expectation is obviously
the squared magnetization.
The model is fully solved at temperature zero \cite{dsgu}, where it exhibits a transition,
which is therefore reproduced at all temperatures along the critical line 
$\beta^{\prime}(\alpha,\beta(\alpha))=1$.
The convexity of the derivative of $A^{\prime}$ means that along the critical line
$\Omega(\sigma_{i_{0}}\sigma_{j_{0}})$ takes values between the values
it takes at $\beta^{\prime}(1/2,\infty)=1$ and at $\beta^{\prime}(\infty,0)=1$,
corresponding to the zero-temperature case and the CW model respectively.
As in both this limiting cases the model is fully solved and 
$\mathbb{E}\Omega(\sigma_{i_{0}}\sigma_{j_{0}})=\langle m^{2}\rangle$ 
has the same behavior
(the same critical index), this means that the critical index is the same
along the whole critical line $\beta^{\prime}(\alpha,\beta(\alpha))=1$.


\section{Fluctuations}

In this section we will compute, in the high temperature region,
the fluctuations of the rescaled magnetization $\mu=\sqrt{N}m$,
which is shown to tend to a centered Gaussian with variance depending on the
connectivity and on the temperature in such a way that it diverges as the critical line
is approached. The method extends to all multi-overlaps, whose variance remains
finite at the critical line. The results are therefore very similar to those found in dilute spin
glasses \cite{gt}. 

The strategy we employed is the one developed in \cite{talaexp}, which was 
then extended in \cite{gt2, gt}. The first step one has to take is the control of the
way the magnetization goes to zero with the size of the system in the high temperature
region, using a suitable perturbation.
The result is that the rescaled magnetization remains finite in the thermodynamic
limit. This is proven in the next subsection. This result is already desirable in itself, 
but our ultimate purpose is to prove that the distribution of the rescaled 
magnetization is Gaussian, and the proof relies on the first result.

\subsection{Bound for the overlaps}

Recall the name we gave the symmetric pressure:
$$
\tilde{A}_{S}(\alpha, \beta)=\ln 2+\alpha\ln\cosh\beta\ .
$$
We want to prove that the squared magnetization vanishes in the thermodynamic
limit not slower than the inverse size of the system.
Let us recall the definition of overlap $q_{1\cdots n}$ among $n$ configurations
$\sigma^{(1)},\ldots ,\sigma^{(n)}$:
$$
q_{1\cdots n}=\frac1N \sum_{i=1}^{N}\sigma^{(1)}_{i}\cdots\sigma^{(n)}_{i}\ .
$$
For $n=1$ one clearly recovers the magnetization $m$ defined in (\ref{magnetization}).
\begin{theorem}
In the high temperature region defined by
$$
2\alpha\tanh\beta<1
$$
the following result holds
$$
\langle q^{2}_{1\cdots n}\rangle=
\mathbb{E}\Omega^{n}(\sigma_{i_{0}}\sigma_{j_{0}})\leq
\mathbb{E}\Omega(\sigma_{i_{0}}\sigma_{j_{0}})=\langle m^{2}\rangle
=O(1/N)\ ,
$$
where $n$ is a natural number larger than one and $m$ is the magnetization
defined in (\ref{magnetization}).
\end{theorem}
Remember that $i_{0},j_{0}$ are independent of the $\{i_{\nu},j_{\nu}\}$
inside $\Omega$, where $\nu\geq 1$.
{\bf Proof}. We will often omit the dependence of the various pressures on
$\alpha$ and $\beta$.

We know from (\ref{alphaderivative}) that
$$
\partial_{\alpha}A_{N}=\partial_{\alpha}\tilde{A}_{S}
+\mathbb{E}\ln(1+\Omega(\sigma_{i_{0}}
\sigma_{j_{0}})\tanh\beta)\leq \ln\cosh\beta
+\langle m^{2}\rangle\tanh\beta
$$
since clearly $\ln(1+x)\leq x$. But we also have 
$\ln(1+x)\geq x\ln 2$ for $0\leq x \leq 1$.
Hence
$$
0\leq \langle m^{2}\rangle\ln 2\tanh\beta
\leq\partial_{\alpha}(A_{N}-\tilde{A}_{S})\leq 
\frac12 \langle m^{2}\rangle 2\tanh\beta\ .
$$
It is clear then that we have to obtain an estimate for 
$\partial_{\alpha}(A_{N}-\tilde{A}_{S})$.
As in \cite{talaexp,gt2,gt}, we will not compare directly 
$A_{N}$ and $\tilde{A}_{S}$. We will
rather compare $\tilde{A}_{S}$ to a perturbed pressure with a larger weight given
to configurations with non-zero magnetization. To the purpose, let us
define, for $\lambda \geq 0$,
$$
\bar{A}(\lambda)=\frac1N \mathbb{E}\ln\sum_{\sigma}\exp(-\beta _{N}(\sigma)
+\lambda N m^{2}/2)
$$
which is convex in $\lambda$ and such that $\bar{A}(0)=A_{N}$. The idea is that if 
the field forcing the magnetization to be strictly positive is not too strong, then
the magnetization will still vanish, provided the temperature is high enough.

Notice that
$$
(\partial_{\lambda}\bar{A})(0)=\frac12 \langle m^{2}\rangle \geq 0\ ,
$$
which by convexity means $\bar{A}(\lambda)\geq \bar{A}(0)=A_{N}$.
Convexity also implies
$$
\frac12 \lambda \langle m^{2}\rangle = \lambda (\partial_{\lambda}\bar{A})(0)
\leq \bar{A}(\lambda)-\bar{A}(0)=\bar{A}(\lambda)-A_{N}\ .
$$
Therefore
$$
0\leq \partial_{\alpha}(A_{N}-\tilde{A}_{S})\leq \frac{2\tanh\beta}{\lambda}
[\bar{A}(\lambda)-A_{N}]\ .
$$
Since $2\alpha\tanh\beta=\beta^{\prime}<1$, we can choose $\lambda_{0}$ such that
$$
\beta^{\prime}=2\alpha\tanh\beta<\lambda_{0}<1\ .
$$
Now let us estimate $\bar{A}(\lambda)$ when $\lambda$ is chosen to depend on
$\alpha$ according to $\lambda=\lambda_{0}-2\alpha\tanh\beta\geq 0$, which
also means $\lambda\leq 1$.
A simple calculation gives
$$
\partial_{\alpha}\bar{A}(\lambda(\alpha))=\ln\cosh\beta+\mathbb{E}
\ln[1+\Omega_{\lambda}(\sigma_{i_{0}}\sigma_{j_{0}})\tanh\beta]
+\frac12 \langle m^{2}\rangle_{\lambda}\frac{d\lambda}{d\alpha}
$$
$$
\leq \ln\cosh\beta+\langle m^{2}\rangle_{\lambda}\tanh\beta-
\langle m^{2}\rangle_{\lambda}\tanh\beta=\ln\cosh\beta\ ,
$$
where the index $\lambda$ reminds that the Boltzmann-Gibbs measure
is now defined in terms of the new weights with the field $\lambda$, due to the 
derivative.
Integrating back against $\alpha$ one then obtains
$$
\bar{A}(\lambda(\alpha))\leq \bar{A}(\lambda(\alpha))|_{\alpha=0}+\alpha\ln\cosh\beta\ ,
$$
with
$$
\bar{A}(\lambda(\alpha))|_{\alpha=0}=
\frac1N \ln\sum_{\sigma}\exp(\lambda_{0}Nm^{2}/2)\ .
$$
Notice that this is the pressure of the CW model at inverse temperature
$\lambda_{0}$, which incidentally is larger than $\beta^{\prime}$
but smaller than one, which is the critical point of the CW model. 
One can thus take advantage of the known properties of
the finite size corrections to the pressure of this model, 
or simply just estimate the right hand side directly using standard techniques.
Let us check that the finite size 
correction to the well known limiting value $\ln 2$ is of order $1/N$:
$$
\bar{A}(\lambda(\alpha))|_{\alpha=0}=\ln 2+O(1/N)\ .
$$
Introducing the centered unit Gaussian variable $J$, we have
\begin{multline*}
\frac1N \ln\sum_{\sigma}\exp(\lambda_{0}Nm^{2}/2)=
\frac1N \ln\sum_{\sigma}\mathbb{E}_{J}\exp(J\sqrt{\lambda_{0}N}m)\\
=\frac1N \ln\mathbb{E}_{J}\prod_{i=1}^{N}
\sum_{\sigma_{i}}\exp(J\sqrt{\lambda_{0}/N}\sigma_{i})=
\frac1N \ln\mathbb{E}_{J}\bigg(2\cosh(J\sqrt{\lambda_{0}/N})\bigg)^{N}\\
=\ln 2+\frac1N\int\frac{dz}{\sqrt{2\pi}}\exp[N\ln\cosh(z\sqrt{\lambda_{0}/N})-z^{2}/2]
\end{multline*}
where $\mathbb{E}_{J}$ is clearly the expectation with respect to the
random variable $J$.
Now we perform the substitution $y=z/\sqrt{N}$ and use the simple inequality
$2\ln\cosh(y\sqrt{\lambda_{0}})\leq y^{2}\lambda_{0}$ for $\lambda_{0}<1$
to get
\begin{multline*}
\frac1N \ln\sum_{\sigma}\exp(\lambda_{0}Nm^{2}/2)\leq \ln 2 +
\frac1N\int\frac{d\sqrt{N}y}{\sqrt{2\pi}}\exp[-N(1-\lambda_{0})y^{2}/2]\\
=\ln 2 +\frac1N\int\frac{dz}{\sqrt{2\pi}\sqrt{1-\lambda_{0}}}\exp[-z^{2}/2]
=\ln 2 + \frac1N \bigg(\frac12\ln\frac{1}{1-\lambda_{0}}+1\bigg)\ ,
\end{multline*}
so that
$$
\frac1N \ln\sum_{\sigma}\exp(\lambda_{0}Nm^{2}/2)\leq \ln 2+O(1/N)\ .
$$
Notice the crucial role played by the choice $\lambda_{0}<1$ here.
So we have now learnt that 
$$
\bar{A}(\lambda(\alpha))\leq \ln 2+\alpha\ln\cosh\beta+O(1/N)=\tilde{A}_{S}+O(1/N)
$$
and in the end
$$
0\leq  \partial_{\alpha}(A_{N}-\tilde{A}_{S})\leq \frac{2\tanh\beta}{\lambda}
(\tilde{A}_{S}-A_{N})+\frac{2\tanh\beta}{\lambda}O(1/N)\ ,
$$
where recall that $\tilde{A}_{S}-A_{N}\leq 0$. Hence it must be
$$
\partial_{\alpha}(A_{N}-\tilde{A}_{S})=O(1/N)\ ,
$$
which proves the theorem since, as we notice at the beginning of the proof,
$\langle m^{2}\rangle\ln 2\tanh\beta
\leq\partial_{\alpha}(A_{N}-\tilde{A}_{S})$. $\Box$

Let us proceed a bit further with a few observations
summarizing our findings about the relative size of $A_{N}$, $\tilde{A}_{S}$,
$\bar{A}(\lambda)$.
As $A_{N}$ and $\tilde{A}_{S}$ share the same value at $\alpha=0$
the fundamental theorem of calculus ensures that
$$
A_{N}-\tilde{A}_{S}=O(1/N)
$$
and we also found that
$$
\bar{A}(\lambda)=\tilde{A}_{S}+O(1/N)=A_{N}+O(1/N),\quad 0\leq\lambda\leq 1\ ,
$$
which means that, as $\tilde{A}_{S}$ does not depend on $\lambda$,
the function $\bar{A}(\lambda)$ has a very small variation, and by convexity
this means that its derivative (at zero, where the derivative is smallest and gives 
$\langle m^{2}\rangle$) is also as small, namely of order $1/N$.

\subsection{Fluctuations}

In this section we find the probability distribution of the rescaled magnetization
$\mu=\sqrt{m}$ in the high temperature region. The result is that such
a probability distribution is a centered Gaussian with variance diverging as
the critical line is approached. A key role in the proof is played by the
finiteness of the rescaled magnetization in the thermodynamic limit 
proven in the previous subsection. As the results of the previous section
hold for all multi-overlaps as well, the results we are about to prove for the 
magnetization also extend to multi-overlaps, although their covariance
does not diverge at the critical line. Not surprisingly, the strategy 
is based on the calculation of the generating function, following again, like in the
previous subsection, the technique developed in \cite{talaexp} and successfully
employed in \cite{gt2,gt}.

\begin{theorem}
Let $\mu=\sqrt{N}m$, where $m$ is the magnetization defined in (\ref{magnetization}). 
Then, as $N\to\infty$, the variable $\mu$ tends 
in distribution to a centered gaussian variable with variance
$$
\langle\mu^{2}\rangle=\frac{1}{1-2\alpha\tanh\beta}\ .
$$
\end{theorem}
{\bf Proof}.
We will employ the standard method, relying on the characteristic generating
function
$$
\phi(u)=\langle \exp(iu\mu)\rangle\ ,
$$
which we will show to be such that
$$
\phi(u)\to\frac{1}{2(1-2\alpha\tanh\beta)}
$$
in the thermodynamic limit.

It is obvious that
$$
\partial_{u}\phi(u)=i\langle\mu\exp(iu\mu)\rangle
=i\sqrt{N}\langle\sigma_{N}\exp(iu\mu)\rangle
$$
by symmetry with respect to permutations of spins.
 
The strategy of the proof consists in estimating the effect of
the removal of the last spin, in the spirit of the cavity method 
\cite{talaexp,gt2,gt,talabook}.
To the purpose, let introduce the notations
$$
u_{-}=u\sqrt{1-1/N}\ , \quad \mu_{-}=\sum_{i=1}^{N-1}\sigma_{i}/\sqrt{N-1}
\ , \quad \alpha_{-}=\alpha(1-1/(N-1))\ .
$$
It is not difficult to check that
$$
\langle\sigma_{N}\exp(iu\mu)\rangle=
\langle\sigma_{N}\exp(iu\sigma_{N}/\sqrt{N}+iu_{-}\mu_{-})\rangle
$$
and as a consequence
$$
\partial_{u}\phi(u)=-u\phi(u)+i\sqrt{N}\langle\sigma_{N}\exp(iu_{-}\mu_{-})\rangle
$$
up to a vanishing term.

We may now assume, with an error of order $1/N$, that 
none of the terms, labelled by $\nu$ and summed up, 
in the Hamiltonian (\ref{hamiltonian}) is $\sigma_{N}\sigma_{N}$, so to have
$$
i\sqrt{N}\langle\sigma_{N}\exp(iu_{-}\mu_{-})\rangle\!=
i\sqrt{N}\mathbb{E}\frac{\Omega_{-}[\exp(iu_{-}\mu_{-})\frac12\!
\sum_{\sigma_{N}}\sigma_{N}\exp(\beta\sigma_{N}\sum_{\nu}^{\kappa}
\sigma_{l_{\nu}})]}{\Omega_{-}[\frac12
\sum_{\sigma_{N}}\sigma_{N}\exp(\beta\sigma_{N}\sum_{\nu}^{\kappa}
\sigma_{l_{\nu}})]}
$$
where $\Omega_{-}$ is the Boltzmann-Gibbs measure associated
with the system with $N-1$ spins, at connectivity $\alpha_{-}$, the variables
$l_{\nu}$ are distributed uniformly over $\{1,\ldots , N-1\}$, and
$\kappa$ is a Poisson random variable of mean $2\alpha$ (see for instance
\cite{lds1} for detailed calculations of this sort). All these quenched random variables
are independent of those appearing in the weights of $\Omega_{-}$.
At this point we proceed following the idea of \cite{talaexp} and define
\begin{eqnarray*}
A & = & \Omega_{-}\left[\exp(iu_{-}\mu_{-})
\sum_{\sigma_{N}}\sigma_{N}\exp(\beta\sigma_{N}\sum_{\nu}^{\kappa}
\sigma_{l_{\nu}})/2\right] \\
B & = & \Omega_{-}\left[\sum_{\sigma_{N}}\sigma_{N}\exp(\beta\sigma_{N}\sum_{\nu}^{\kappa}
\sigma_{l_{\nu}})/2\right]\\
\tilde{B} & = & \cosh^{\kappa}\beta
\end{eqnarray*}
which will be used in the following trivial identity
$$
\frac{A}{B}=2\frac{A}{\tilde{B}}-\frac{AB}{\tilde{B}^{2}}
+\frac{A}{B}\left(1-\frac{B}{\tilde{B}}\right)\ .
$$
From the identity above is should be clear that the key idea 
is to simplify the denominator $B$, dealing with $\tilde{B}$ instead.
So we have three terms to compute. We want to show that the first two
give the same result, and the third is negligible in the thermodynamic limit.

Let us define $\mathbb{E}_{\kappa}$ as the expectation with respect to $\kappa$
and $\mathbb{E}_{l}=\prod_{\nu=1}^{\kappa}\mathbb{E}_{l_{\nu}}$ is the
expectation with respect to the random sites appearing explicitly in $A$ and $B$ only;
the other quenched variables implicitly included in $\Omega_{-}$ are excluded by these
expectations.

Let us start from $A$, whose core is
\begin{multline*}
\mathbb{E}_{\kappa}\mathbb{E}_{l}\sum_{\sigma_{N}}
\sigma_{N}\exp(\beta\sigma_{N}\sum_{\nu}^{\kappa}\sigma_{l_{\nu}}) 
 = 
\mathbb{E}_{\kappa}\mathbb{E}_{l}\sum_{\sigma_{N}}\sigma_{N}
\prod_{\nu=1}^{\kappa}\exp(\beta\sigma_{N}\sigma_{l_{\nu}}) \\
 =  \mathbb{E}_{\kappa}2\cosh^{\kappa}\beta\sum_{\sigma_{N}}
\sigma_{N}\prod_{\nu=1}^{\kappa}\mathbb{E}_{l_{\nu}}(1
+\sigma_{N}\sigma_{l_{\nu}}\tanh\beta)\\
 = \mathbb{E}_{\kappa}2\tilde{B}\sum_{\sigma_{N}}\sigma_{N}(1
+\sigma_{N}\mu_{-}/\sqrt{N}\tanh\beta)^{\kappa}\ .
\end{multline*}
Now $\tilde{B}$ clearly cancels out in the fraction $A/\tilde{B}$ we are computing.
At this point the formula in (\ref{exponential}) is employed and lets us obtain
\begin{eqnarray*}
\frac{A}{\tilde{B}} & = & =\Omega_{-}[\exp(iu_{-}\mu_{-})\frac12\sum_{\sigma_{N}}
\sigma_{N}\exp(2\alpha\tanh\beta\sigma_{N}\mu_{-}/\sqrt{N-1})] \\
{} & = & \Omega_{-}[\exp(iu_{-}\mu_{-})\sinh
(\mu_{-}2\alpha\tanh\beta/\sqrt{N-1})]\ .
\end{eqnarray*}
Now this is where the result of the previous section 
$$
\sup_{N}\langle\mu^{2}\rangle<\infty
$$
becomes crucial, as it ensures that in the 
thermodynamic limit the $\sinh$ can be replaced by its
first order approximation, i.e. its argument, so that when $N\to\infty$ we have
$$
i\sqrt{N}\mathbb{E}\frac{A}{\tilde{B}}=2\alpha\tanh\beta
\langle i\mu\exp(iu\mu)\rangle=(2\alpha\tanh\beta)\partial_{u}\langle\exp(iu\mu)\rangle\ .
$$

Let us proceed with the slightly more involved term $AB/\tilde{B}$, focussing on
the numerator first
\begin{multline*}
\mathbb{E}_{\kappa}\mathbb{E}_{l}AB=\Omega_{-}\left[e^{iu_{-}\mu_{-}}
\mathbb{E}_{\kappa}\mathbb{E}_{l}\frac14\sum_{\sigma_{N},\sigma^{\prime}_{N}}
\sigma_{N}e^{\beta\sigma_{N}\sum_{\nu}^{\kappa}\sigma_{l_{\nu}}}
e^{\beta\sigma^{\prime}_{N}\sum_{\nu}^{\kappa}\sigma^{\prime}_{l_{\nu}}}\right]=\\
\mathbb{E}_{\kappa}\tilde{B}^{2}\Omega_{-}\left[e^{iu_{-}\mu_{-}}
\frac14\sum_{\sigma_{N},\sigma^{\prime}_{N}}\sigma_{N}\prod_{\nu=1}^{\kappa}
\mathbb{E}_{l_{\nu}}(1+\sigma_{N}\sigma_{l_{\nu}}\tanh\beta)
(1+\sigma^{\prime}_{N}\sigma^{\prime}_{l_{\nu}}\tanh\beta)\right]\ ,
\end{multline*}
where we used $\sigma^{\prime}$ as a second label for configurations, since
in the product $AB$ there are two summations over the spin configurations.
Again, $\tilde{B}^{2}$ is removed by the denominator and therefore
\begin{multline*}
\mathbb{E}\frac{AB}{\tilde{B}^{2}}=\langle \exp(iu_{-}\mu_{-})
\frac14\sum_{\sigma_{N},\sigma^{\prime}_{N}}\sigma_{N}\times\\
\mathbb{E}_{\kappa}\prod_{\nu=1}^{\kappa}
\mathbb{E}_{l_{\nu}}(1+\sigma_{N}\sigma_{l_{\nu}}\tanh\beta+
\sigma^{\prime}_{N}\sigma^{\prime}_{l_{\nu}}\tanh\beta+
\sigma_{N}\sigma_{l_{\nu}}
\sigma^{\prime}_{N}\sigma^{\prime}_{l_{\nu}}\tanh^{2}\beta)
\end{multline*}
\begin{multline*}
=\langle \exp(iu_{-}\mu_{-})
\frac14\sum_{\sigma_{N},\sigma^{\prime}_{N}}\sigma_{N}\times\\
\mathbb{E}_{\kappa}(1+\sigma_{N}m_{-}\tanh\beta+
\sigma^{\prime}_{N}m^{\prime}_{-}\tanh\beta+
\sigma_{N}\sigma^{\prime}_{N}q_{-}\tanh^{2}\beta)^{\kappa}\rangle
\end{multline*}
where $m_{-}=\mu_{-}/\sqrt{N-1}$ is the magnetization of the first $N-1$ spins
and $q_{-}$ is the overlap between two configurations of the first $N-1$ spins.
Using again (\ref{exponential}) we get
\begin{multline*}
\mathbb{E}\frac{AB}{\tilde{B}^{2}}=\langle  \exp(iu_{-}\mu_{-}) 
\frac14\sum_{\sigma_{N},\sigma^{\prime}_{N}}\sigma_{N}\times\\
\exp[2\alpha(\sigma_{N}m_{-}\tanh\beta+
\sigma^{\prime}_{N}m^{\prime}_{-}\tanh\beta+
\sigma_{N}\sigma^{\prime}_{N}q_{-}\tanh^{2}\beta)]\rangle
\end{multline*}
\begin{multline*}
=\langle  \exp(iu_{-}\mu_{-}) 
\frac12\sum_{\sigma^{\prime}_{N}}\exp(2\alpha\sigma^{\prime}_{N}
m^{\prime}_{-}\tanh\beta)\times\\
\frac12\sum_{\sigma_{N}}\sigma_{N}
\exp[2\alpha(\sigma_{N}(m_{-}\tanh\beta+\sigma^{\prime}_{N}
q_{-}\tanh^{2}\beta))]\rangle
\end{multline*}
\begin{multline*}
=\langle  \exp(iu_{-}\mu_{-}) 
\frac12\sum_{\sigma^{\prime}_{N}}\exp(2\alpha\sigma^{\prime}_{N}
m^{\prime}_{-}\tanh\beta)\times\\
\sinh[2\alpha(m_{-}\tanh\beta+
\sigma^{\prime}_{N}q_{-}\tanh^{2}\beta)]\rangle
\end{multline*}
and again keeping only the terms surviving in the thermodynamic limit
we can write
\begin{multline*}
\mathbb{E}\frac{AB}{\tilde{B}^{2}}=\langle  \exp(iu_{-}\mu_{-}) 
\frac12\sum_{\sigma^{\prime}_{N}}\exp(2\alpha\sigma^{\prime}_{N}
m^{\prime}_{-}\tanh\beta)
m_{-}\rangle2\alpha\tanh\beta\\
+\langle  \exp(iu_{-}\mu_{-}) 
\frac12\sum_{\sigma^{\prime}_{N}}\exp(2\alpha\sigma^{\prime}_{N}
m^{\prime}_{-}\tanh\beta)
\sigma^{\prime}_{N}q_{-}\rangle2\alpha\tanh^{2}\beta
\end{multline*}
\begin{multline*}
=\langle  \exp(iu_{-}\mu_{-}) \cosh(2\alpha m^{\prime}_{-}\tanh\beta)
m_{-}\rangle2\alpha\tanh\beta\\
+\langle  \exp(iu_{-}\mu_{-}) 
\sinh(2\alpha m^{\prime}_{-}\tanh\beta)q_{-}\rangle2\alpha\tanh^{2}\beta
\end{multline*}
which in the limit reduces to
\begin{multline*}
i\sqrt{N}\mathbb{E}\frac{AB}{\tilde{B}^{2}}=
i\sqrt{N}\langle  \exp(iu\mu)m\rangle2\alpha\tanh\beta+\\
i\sqrt{N}\langle  \exp(iu_{-}\mu_{-}) 
 m^{\prime} q_{-}\rangle 4\alpha^{2}\tanh^{3}\beta\\
\to i\langle  \exp(iu\mu)\mu\rangle2\alpha\tanh\beta
=(2\alpha\tanh\beta)\partial_{u}\phi(u)
\end{multline*}
as $\cosh(2\alpha m^{\prime}\tanh\beta)\to 1$ and
$\sqrt{N}\langle m^{\prime} q_{-}\rangle\to 0$.

The last term we have to consider is $B/\tilde{B}$:
$$
\mathbb{E}\frac{B}{\tilde{B}}=\mathbb{E}\Omega_{-}
[\frac{1}{2\tilde{B}}\sum_{\sigma_{N}}\exp(\beta\sigma_{N}\sum_{\nu}\sigma_{l_{\nu}})]
=\mathbb{E}\Omega_{-}[\frac{1}{2}\sum_{\sigma_{N}}\mathbb{E}_{\kappa}
(1+\sigma_{N}m_{-}\tanh\beta)
$$
$$
=\mathbb{E}\Omega_{-}[\frac{1}{2}\sum_{\sigma_{N}}\exp(\sigma_{N}2\alpha\tanh\beta)]
=\mathbb{E}\Omega_{-}\cosh(m_{-}2\alpha\tanh\beta)\to1\ .
$$
Taking into account the orders of magnitude as before
$$
\sqrt{N}\mathbb{E}\exp(2\beta\kappa)\left(1-\frac{B}{\tilde{B}}\right)^{2}\to 0
$$
and
$$
\sqrt{N}|\mathbb{E}\frac{A}{B}(1-\frac{B}{\tilde{B}})^{2}|\leq \sqrt{N}
\exp(2\beta\alpha)\mathbb{E}\left(1-\frac{B}{\tilde{B}}\right)^{2}
$$
since $B\geq 1$ by Jensen inequality
and 
$$
|A|\leq \Omega|\exp(\beta\sigma_{N}\sum_{\nu}\sigma_{l_{\nu}})|
\leq \Omega(\exp(\beta\kappa))\leq \exp(\beta \kappa)\ .
$$
Thus this last third term does not contribute in the limit and 
the final result is
$$
(1-2\alpha\tanh\beta)\partial_{u}\phi_{u}=-u\phi(u)
$$
which completes the proof recalling that $\phi(0)=1$. $\Box$

A straightforward generalization of the previous theorem, which requires 
conceptually similar but longer calculations, provides the next 
\begin{theorem}
Let $\mu_{1\cdots n}=\sqrt{N}q_{1\cdots n}$, where $q_{1\cdots n}$
is the overlaps between $n$ configurations. Then
\begin{eqnarray*}
&& \langle\mu^{2}_{1\cdots n}\rangle  =  \frac{1}{1-2\alpha\tanh^{n}\beta}\ , \\
&& \langle\mu_{1\cdots n}\mu_{k\cdots k+n}\rangle=0
\qquad\mbox{if} \quad k\neq 0\ , \\
&& \langle\mu_{1\cdots n}\mu_{1\cdots m}\rangle=0  
\qquad\mbox{if} \quad n\neq m\ ,
\end{eqnarray*}
in the thermodynamic limit.
\end{theorem}


\section{Outlook}

The computation of the free energy at high temperature
and at zero temperature has been followed by progresses
in two directions. The complete characterization (in the present work)
of the statistical properties of the model 
in the same high temperature region on the one hand,
the rigorous computation of the free energy
at any temperature on the other hand \cite{dm}.
The next important step is then a good understanding of
the low temperature phase, where there are  
unanswered questions and interesting hints \cite{gparisi}.


\section*{Acknowledgments}

The author gratefully thanks Francesco Guerra, with whom the research
about this model was started and supported through so many inspiring discussions.

Work supported by European Commission Contract
FP6-2004-NEST-PATH-043434 (CULTAPTATION).


\end{document}